\documentclass[prl,showpacs,showkeys,groupedaddress]{revtex4} 

\usepackage{color}

\definecolor{blue}{rgb}{0.00,0.00,1.00}
\definecolor{green}{rgb}{0.00,1.00,0.00}
\definecolor{red}{rgb}{1.00,0.00,0.00}
\definecolor{purple}{rgb}{0.63,0.13,0.94}
\definecolor{yellow}{rgb}{1.00,1.00,0.00}

\usepackage{graphicx}

\begin{document}


\title{Electromagnetic Composites at the Compton Scale}

\author{Frederick J. Mayer}
\affiliation{Mayer Applied Research Inc., 1417 Dicken Drive, Ann Arbor, MI
48103, email: fmayer@sysmatrix.net}

\author{John R. Reitz}
\affiliation{  2260 Chaucer Court, Ann Arbor, MI 48103
email: jrreitz@ameritech.net}

\date{\today}

\begin{abstract}
A new class of electromagnetic composite particles is proposed. The 
composites are very small (the Compton scale), potentially long-lived, 
would have unique interactions with atomic and nuclear systems, and, if they exist, could 
    explain a number of otherwise anomalous and conflicting observations in diverse 
    research areas.

\end{abstract}

\maketitle

\section{I. Introduction}
\par In recent years there have been a number of experimental observations that 
     are difficult to explain within our now-standard models of 
atomic and nuclear physics and cosmology. The case of so-called 
``dark matter'' is an example. It appears that only a  
small fraction of the mass of the universe is constructed from ordinary protons, neutrons, and electrons. 
So, many cosmologists have turned to some relic elementary particle as the candidate to complete the 
mass deficit. Strange observations such as the excess heat from 
the earth and ``cold fusion'' are still other examples. 
We have wondered if there might be configurations of nucleons and electrons 
that would not be directly observable in the same 
way as are the ordinary nucleon atomic systems. This consideration was the genesis of the 
work presented here.
\par The possibility of new electromagnetic bound states in which the magnetic and electric
	forces are treated equally and are of comparable size was suggested in our recent 
	paper \cite{ref1}. For example, the electrostatic force between two electrons 
	${e^{2}}/{r^{2}}$  
	is comparable with the dipole-dipole magnetic force ${\mu_{e}^{2}}/{r^{4}}$  at 
	a distance ${r}\,{\approx}\,{\lambda_{c}}$, where $\lambda_{c}$   
	is the electron Compton wavelength. In fact, a number of bound states involving 
	two electron-like particles were found as solutions to the Dirac equation. 
	However, none of these states involved nucleons because the nuclear magnetic 
	moments are too small to produce binding.  Yet, it seemed plausible that composites 
	that included nucleons might be possible at the Compton scale. 
	These composites might resemble normal atoms perhaps with different 
	characteristics, but would be, of course, much smaller than atoms. 
	\par In this paper, we propose simple composite systems that include 
	nucleons but are still bound together by comparable electric and magnetic forces.  
	These entities make up a three-body system which is too complicated to 
	treat rigorously in a  
	quantum mechanical manner, so we present a simple Schr\"{o}dinger model (one which is consistent with its Dirac equation origin) to get quantitative 
	estimates of the system's size and binding energy. Clearly, without a 
	quantum electrodynamical formulation for these composites, their existence 
	is unproven; however, since these entities appear plausible, we will look at the 
	consequences as if they do exist.
	\par We first describe several model calculations for these 
	three-body systems and determine whether bound states appear possible.
	Second, we examine the situations in which these composites might 
	be expected to be formed. Finally, we connect the characteristics of the 
	proposed composite particles to a number of anomalous observations over 
	the past years. In later papers, we will consider some of these anomalous observations in detail.

\section{II. The Electromagnetic Configuration}
\par	The simplest classical model of one of these three-body systems consists of a positively 
charged nucleus ($Ze$) and two ``point" electrons on opposite sides of the nucleon.  The electrons have their
customary magnetic moment $\mu_{e}=e\lambda_{c}/2$.  The nucleon provides the attractive electrostatic
force pulling the electrons together; the electrons repel each other electrostatically
and magnetically through the dipole-dipole interaction.  The electronic motion must be
highly correlated -- i.e., the electrons move in such a way that they stay apart as far as possible,
consistent with maximizing their interaction with the nucleon. Is this a possible configuration at the Compton
scale, i.e., is the electron magnetic moment developed to the point that  $\bf{\mu_{e}} \cdot \bf{B}$
represents an appropriate energy term?  There is no such term in the Dirac equation, only a vector
potential that interacts with the electronic motion.  However, 
Schiff \cite{ref2} shows that the Dirac equation for an electron in an electromagnetic field is equivalent to a Schr\"{o}dinger equation
with the usual $\bf{\mu_{e}} \cdot \bf{B}$ term [Schiff's Eq. (43.27)] if
$2mc^{2}\gg  E' - e\phi$ where $E'=E_{total}-mc^{2}$. This turns out to be the case for
the model considered herein.  
\par Of course, one aspect of the electronic motion predicted by the Dirac
equation does not appear in the Schr\"{o}dinger equation, namely, the
\emph {zitterbewegung} \cite{ref3}.  This motion occurs on a very short time scale, is
confined to small distances (of the order of $\lambda_{c}$), does not affect the classical
trajectory, and is believed to be responsible for creating the electron-spin
magnetic moment. However, the \emph {zitterbewegung} must traverse distances 
of at least one Compton in order to develop the observed magnetic 
moment. The classical models described below have equilibrium radii 
smaller than this, so the magnetic moments cannot be treated as point 
entities in any final picture.

\par There are two cases with no orbital angular momentum. 
In the first case, the electrons are located equatorially, on opposite sides, at distance 
$r$ from a nucleon $Ze$ - see Fig.1(a).

\begin{figure}[htbp]
\centering
\resizebox{8cm}{6cm}
{\includegraphics{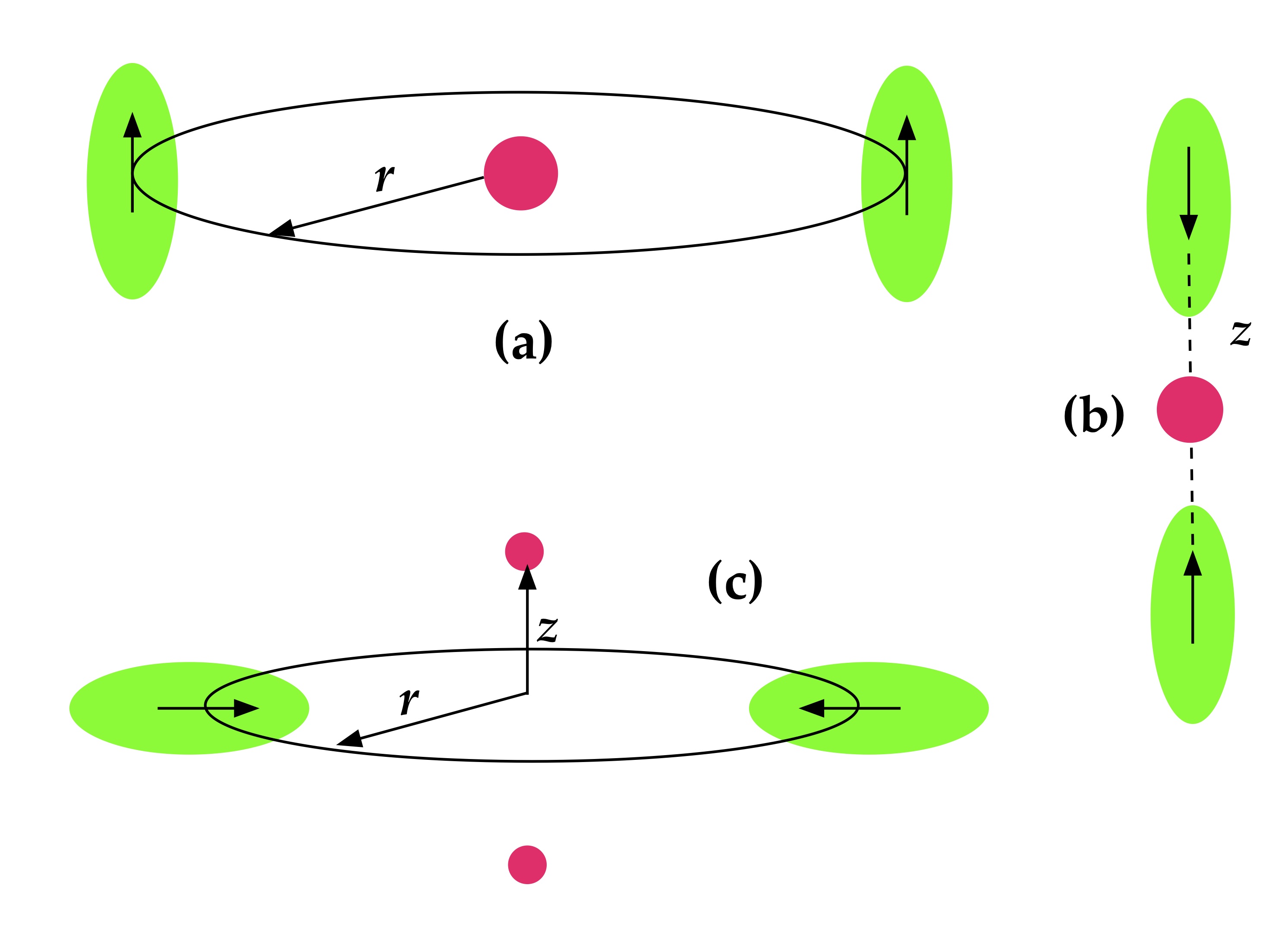}} 
\caption{Compton composite ``classical" depictions showing {\color{red}{protons}}  and  {\color{green}{electrons}}.}
\end{figure}
 
 We neglect the nuclear magnetic moment. 
Consider the case where the total angular momemtum is 
zero, i.e., when one electron's particle momentum is canceled by the 
magnetic field momentum of the other electron.
\begin{equation}
	n \hbar  =r (m v   + (e/c) A_{\phi}) = m v r - (e/c)\mu_{e}/ 4r = 0 
\end{equation}
where $\mu_{e}$ is the electron's magnetic moment.  The centripetal force 
equation for one of the electrons is then
\begin{equation}
m v^{2}/ r  = Z_{eff} e^{2}/r^{2}  -  (e/c) \mu_{e}v/8r^{3}  -  3\mu_{e}^{2}/16r^{4}
\end{equation} where $Z_{eff}=Z-1/4$.
Equation (1) gives   
\begin{equation}
\beta =  v/c  = \alpha/ 8 r^{2}
\end{equation} where $\alpha$ is the fine structure constant, and (\bf{NOTE}\rm) from here 
and henceforth, distances are measured in units of $\lambda_{c}$, 
and energies in units of 
\mbox{$E_{c} = e^{2}/{\lambda_{c}} = 3.7\, \mathrm{{keV}}$}.  These equations will show 
that $\beta \ll 1$. The total electromagnetic energy of this system is  
\begin{equation}
W_{\mathrm{em}}=-2\,Z_{eff}/r +1/(32\,r^{3}).
\end{equation}
Solving Eqs.\,(1) and (2)  gives $r =\sqrt{3/Z_{eff}}/8$. This system has a total electron spin of 1. 
With $Z = 1$, $r=1/4$, and the binding energy $E_{\mathrm{B}} = 4$. 
\par In the second case -- see Fig.1(b), the electrons are located on 
the $z$-axis at $+z$ and $-z$, with the nucleon at the origin.  
The total electromagnetic energy of the system as a function of $z$ 
(the electron-nucleon distance) is
\begin{equation}
W_{\mathrm{em}}(z)  = -2Z_{\mathrm{eff}}/z  +1/16z^{3}.	
\end{equation}
This has a minimum at  $z = \sqrt{6/Z_{\mathrm{eff}}}/8 $;\, for \, $Z  =  
1$, \,  $z = 0.354$.  
We note that this represents a potential which is about 10 keV deep at 
an electron-nucleon distance of about one third of a Compton.
In addition to $W_{\mathrm{em}}$ there is also kinetic energy, presumably due to vibration.

\par	Summarizing, we have found two (classical) bound states for 
three-body electromagnetic composites: a compact equatorial state with spin 1 and a binding energy of about 15 
keV (for $Z=1$), and a more loosely-bound axial state with spin zero. 
But the classical models cannot provide a valid picture. The 
deBroglie wavelength is not short enough to localize an electron in a 
fraction of a Compton.

\par What is required is a non-perturbative QED treatment of the 
three-body system, but this is not presently available. We can, 
however, solve a simplified Schr\"{o}dinger model. Here, again, we 
note that the deBroglie wavelength is not short enough to keep the 
electrons apart and at the same time localize them in the Compton range. 
Thus, the electron wave-functions must overlap, so that an $S=0$ spin 
state is required.
\par  The \emph {zitterbewegung} loops can form around any axis; 
there is no net spin and hence no preferred axis. In writing a 
Schr\"{o}dinger Hamiltonian for one of the electrons we choose a 
Hartree model \cite{ref4} in which the electron under consideration is 
a ``point''-electron \, $e_{1}$ interacting with a nucleus and the 
electron  distribution of $e_{2}$ (electron-2). The Hartree field 
model is a central-field model, so we can use spherical coordinates. We make two assumptions 
to differentiate this new type of wave-function from the atomic case: 
the magnetic interaction between the electrons plays an essential 
role, and the electrons are completely correlated so that $r_{1}$ 
equals $-r_{2}$ at any instant. Since the two electrons in this model 
have the same wave-function we can just as easily solve a Schr\"{o}dinger 
equation for the 2-\emph{electron system}; this takes the form:

\begin{equation}
\left[\frac{\partial^{2}}{\partial{r^{2}}}+\alpha[E-V(r)]\right]\psi(r)=0
 \end{equation}
    where $\alpha$ is the fine structure constant, $E$ is the 
    eigenvalue (the particle binding energy is $-E$), and the 
    potential $V(r)$ is given by:
   \begin{equation}
V(r)=\frac{1}{4\,{\left( 1 + 4\,r^2 \right) }^{\frac{3}{2}}} + 
  \frac{1}{{\sqrt{1 + 4\,r^2}}} - 
  \frac{2\,Z}{{\sqrt{r^2\,+{{r_n}}^2}}}
\end{equation}
where ${r_n}$ is the nuclear radius in Compton units.
\par The first term in Eq.(7)
is the magnetic interaction between the electrons. To calculate this 
we need the axial magnetic field of a loop, and this can be obtained 
from expressions derived by Smythe \cite{ref5}. The spin dipole is produced by 
swirling \emph{zitterbewegung} currents with dimensions of order of 
$\lambda_{c}$, this swirl of currents around the image point of $r_{1}$ 
forms the electron distribution of electron $e_{2}$.
\par  The result of our quantum mechanical calculation is as follows: the two
one-electron wavefunctions overlap and have strong maxima at the origin (the
nucleus). Taking $Z=1$, the wavefunction 
$\psi$\,(squared) is shown in Fig. 2 where the average value of $r$ is about six Comptons and the particle binding
energy $E_{B}\approx 1$ or about 3.7 keV.

\begin{figure}[htbp]
 \centering
 \resizebox{9cm}{6cm}{
 \includegraphics{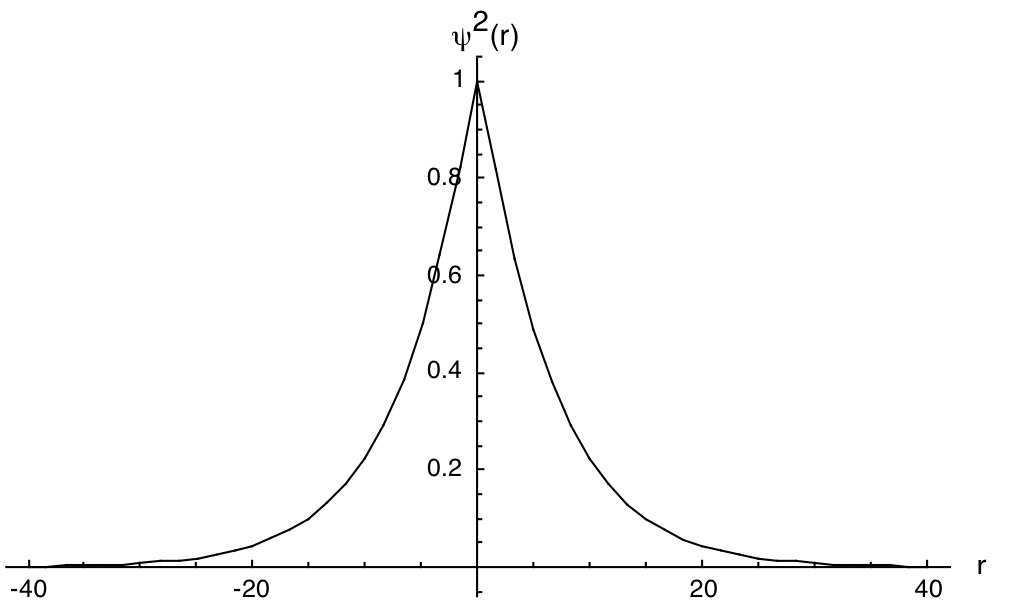} }
 \caption{Plot of $\psi^{2}(r)$ based on a simplified Schr\"{o}dinger model.}
 \end{figure}

\section{III. Characteristics of Hydrogen Tresinos}
\par	For convenience, we have given these three-body electromagnetic composites the name \emph{tresinos}. 
In this section we discuss the properties of the hydrogen tresinos (the $Z=1$ case). Where 
and under what conditions might they be formed and will they survive for 
some length of time? Clearly, the physical conditions favoring tresino 
formation require a nucleon and a source of 
electrons such that two opposing-spin electrons have a reasonable probability by ``falling into" the 
potential well of the nucleon (perhaps requiring another nearby particle to conserve energy and momentum). Sufficiently high density plasmas, either gaseous or metallic,
might be expected to present these conditions. On the other hand, 
ionic materials, in 
which a pair of donor electrons are in close proximity to each 
other (such as might be found in a chemical bond) 
might similarly be advantageous to tresino formation. 
It would appear, however, that tresinos are not
particularly easy to make. Their formation involves the interaction of three particles: protons 
(or deuterons or tritons) and two electrons, in a restricted geometry. A situation 
involving high densities of both protons and electrons would seem to be 
required for them to be produced.

\par Aside from being charged and thus responsive to an electric field, the 
tresino would appear to have little or no interaction with atomic systems. It would, therefore, stay around
until being eventually destroyed or neutralized, very likely through 
attachment to a positive nucleon. The most likely nucleon is another hydrogen 
nucleus ($p$,\, $d$,\, or $t$) and, at least classically, it appears that this 
would be energetically favorable. The attachment would form either a tresino-proton pair (somewhat like a molecule)  or what we will call a ``quatrino''.
\par A classical model of the quatrino is shown in Fig.1(c). It is a 
four-body composite with two hydrogen nuclei located on the axis at 
$\pm z$ and two electrons on a circle of radius $r$ in the midplane. 
The orbital velocity of the electrons is very small, and as before, and there is zero 
angular momentum (see Section II). Using the same type of 
classical analysis used in Section II, we find that $r\,=\sqrt{3}\,z = 
0.211$, and the binding energy of the quatrino is about 25 keV. 
However, we admit that we do not have a quantum mechanical model of 
this complex composite.
\par Why haven't tresinos been seen? As already mentioned, they would not be
readily created, they would not be reactive with electrons or atomic 
systems, and, although the $h$-tresinos are charged, they probably would not 
remain un-neutralized for very long.  However, a tresino-proton pair (or a quatrino) being neutral would be expected to move easily through macroscopic systems. 
The tresinos do not appear to have excited states therefore they would not have photon interactions other than perhaps through rotational or vibrational excitations.
\par Compton composite formation would release some energy (the binding 
energy--3.7 keV for the $h$-tresino from our Schr\"{o}dinger solution). We speculate that 
this heat of formation may have been observed, but misinterpreted, in 
observations discussed in Sections V and VII. In the 1990s 
there were many cases where this heat of formation may have 
been observed and measured, starting with the ``cold fusion'' 
experiments of Fleischmann and Pons \cite{ref6}. Although usually these 
experiments involved deuterium-loaded Pd and/or ionic solutions 
containing deuterium ions, some cases \cite{ref7,ref8} used ordinary hydrogen 
in place of deuterium. The experimenters generally attributed the 
observation of the excess heat to nuclear reactions. But this 
interpretation has not been accepted by nuclear physicists. Still, the 
source of this excess heat, which is more than an order of magnitude 
larger than that from known chemical reactions, has not been 
definitively identified. We propose that much of this heat (perhaps 
all of it in experiments using ordinary ``light'' water) comes from tresino 
formation energy. We present more discussion of these controversial 
observations in Section VII.  

\section{IV. Compton Composites and Dark Matter}
\par	So far, we have been describing the characteristics and interactions 
of the $h$-tresinos which carry a net charge of minus one. Let us now
consider the He-tresino.  As with its atomic counterpart, the He-tresino 
is very strongly bound ($E_{B}=14.3\,\,\mathrm{keV}$) 
and a neutral composite. This particle would be very small, neutral, and have a mass of about 3.7 GeV.
It would be expected to have very few interactions of any kind with ordinary matter and 
would not be ionized except in the cores of very hot stars. As such, it might be a candidate for the 
so-called dark matter in the cosmological context. 
\par The quatrino, if it exists, is also a neutral composite with a 
mass of 1.8 GeV. Here, we are considering the $h$-quatrino made of 
protons. This Compton composite  should  be stable and long-lived, and could also be a dark 
matter candidate.  And yet another possibility for the dark matter: a combination of 
$h$-tresinos and protons, possibly bound together as proton-tresino molecules.
\par	The He-tresino and the $h$-quatrino would be classified as weakly 
interacting massive particles (WIMPs). Interestingly, 
Peacock \cite{ref9}, notes that, ``the universe  may be closed by massive neutrino-like particles
with masses around 3 GeV." But he also states ``that none of the known neutrinos can be as 
massive as 3 GeV." The proton-tresino molecule (or the He-tresino) could have been formed in the early 
universe before the cooling and recombination of ordinary matter and might 
have continued to drift along with the universal expansion, being 
affected only by gravitational forces. These ideas are discussed more fully in a later paper.

\section{V. Compton Composites and Heat from the Earth}
\par 	For some time, it has been realized that there is a substantial discrepancy regarding 
the earth's internal heat source. That there is such a source is not in dispute, but the 
conventional explanation for the earth's internal heat (alpha decay of 
uranium and thorium) has an associated problem. Namely, where is all of the 
helium? At the elevated temperatures of the earth's interior, helium should readily 
escape. Therefore, measurements of the helium effluxing from the earth 
would be expected to be in balance with the radioactive decay heat from 
these nuclides. Yet, this appears not to be the case.
According to some measurements \cite{ref10}, there is approximately 
twenty times more heat than can be accounted for by the helium measured. 
\par	Perhaps tresino formation as mentioned above is 
possible in the high-temperature and pressure materials in the earth.  
There is also water in the earth's crust and mantle, and these conditions may favor 
the formation of Compton composites. If tresino formation energy is, in fact,
the largest source of the earth's heat generation, it would explain why the major source
of this heat comes from the crust and upper mantle, and it would also resolve a number of 
other unexplained anomalies concerning the earth's heat and helium emanations (Mayer and Reitz \cite{ref11}).

\section{VI. Compton Composite-Induced Nuclear Reactions}
\par As discussed in Section III, the $h$-tresino in a hydrogen 
environment ($p$,\,$d$,\,$t$) will probably end up in the vicinity of 
hydrogen ions. Even if there is not permanent 
attachment to the ion, the electron shielding in the tresino will allow frequent 
nuclear encounters at a distance of a Compton or less. This opens up the 
possibility of nuclear reactions.
\par A tresino diffusing through a metal like Pd which has absorbed 
hydrogen (or deuterium) will be attracted to a hydrogen nucleon. 
At sub-coulomb barrier energies,  neutron ``stripping'' 
(or transfer) reactions \cite{ref12} are the most common nuclear reactions. 
Tresino-induced neutron stripping may occur with a 
$d$-tresino operating in an environment containing other 
deuterons (more about these reactions can be found in future paper \cite{ref11}). The electrostatic 
force between the \mbox{$d$-tresino} and a deuteron favors bringing them into close 
proximity and with dynamic electron shielding the two
nuclei may be brought to within a fraction of a Compton. Since the neutron in the deuteron is rather loosely bound 
and the energy for the reaction is favorable, the neutron may be 
picked up by the \mbox{$d$-tresino} (or, in some cases, the pick-up may be in 
the reverse direction) in the reaction $d + d^{*} 
\rightarrow p+t+2\,e + 4\,\mathrm{MeV}$ where we use $d^{*}$ to indicate 
the \mbox{$d$-tresino} and we might expect that the tresino breaks up in the 
process (however, see \cite{ref11}). This type of neutron transfer reaction appears 
to be considerably more probable than 
\emph {compound nucleus} formation requiring much closer encounters.

\section{VII. Compton Composites and ``Cold Fusion"}
\par	The area known as ``cold fusion" \cite{ref13} has received much attention, both good and bad, 
over the past decade. We will not attempt to explain all the 
claims or even all the experimental observations from this complex and  
muddled research area, but we will show that a number of otherwise unexplained observations 
are consistent with tresino induced reactions. A good overview of all of the anomalous
results from this area can be found in a review paper by Storms \cite{ref14}. In addition, we note that there have been a number of models that have sought to find compact, charge neutral, electron-proton systems to explain enhanced screening in ``cold fusion" experiments. In particular, the work of Rice, et al.\,\cite{ref15} examined this issue and concluded that ``models in which the electron is tightly bound to the hydrogen or deuterium nucleus were found to have serious qualitative or quantitative defects". In contrast to their work, the present paper requires two electrons interacting with a proton and must include the electrons' dipole-dipole interaction. Hence, ours is a  quite different Hamiltonian---one that yields the compact (and energetic) bound states presented above.
\newline
\emph{a.  Observations of Excess Heat}
\par  The original papers by Fleishmann and Pons \cite{ref6} introduced the cold 
fusion idea and claimed nuclear fusion as the source of ``excess'' heat in their 
electrochemical (deuterium-loaded Pd) cells. Although the heat was present, 
the expected energetic nuclear reaction products were not. These experiments were repeated by others, including some using non-deuterated water, many reporting ``excess heat".
\par
Now, if $d$-tresinos are formed during deuteron loading of palladium, there 
are at least two possibilities to generate heat. First, there is the binding 
energy of the tresino which is released during its formation ($\approx 
2\times 10^{8}$ joules/gram).  Second, there is the much larger energy per 
reaction if neutron transfer reactions take place. There may have 
been many instances in which the heat of tresino formation has been observed 
but misinterpreted as chemical reaction heat.
\newline 
\emph{b. Observations of Tritium}
\par	For many years, there have been observations of tritium being produced in 
deuterium loaded metal experiments \cite{ref16,ref17}. These observations were not accompanied by 
other nuclear reaction products such as neutrons which might have been 
expected from ordinary $d-d$ fusion reactions because the neutron and 
triton branches, through compound nucleus formation, are about equally probable. Although the tritium 
was many orders of magnitude above background, extensive measurements were made to eliminate the 
possibility of tritium being somehow introduced into the experiment as an impurity.  These experiments may be
explained as the result of the tresino-induced neutron transfer 
reactions in deuterated material (see previous Section). 
\par	Many (but not all) of the controversial claims and observations of this experimental 
	area may have straightforward explanations through the heats of 
	formation of tresinos and quatrinos or through nuclear reactions in which 
	they play a role. This possibility will be the focus of a future paper. 
\section{VIII. Discussion}
We have proposed the existence of a new class of subatomic, 
composite particles which might have eluded direct observation. Although we are unable to present a formal
quantum electrodynamical solution for the Compton composite particles, we have shown 
that their existence is not in conflict with well-established quantum mechanical principles.
\par But perhaps more interesting is the indirect evidence. The 
existence of these particles can provide explanations for a number of 
physical observations which have so far eluded attempts at 
explanation. These include (1) the discrepancy between the heat 
emanating from the earth and its proposed source from radioactive 
material, (2) the unexplained excess heat generated in ``cold 
fusion'' experiments, and (3) the composition of the dark matter 
filling the universe. Perhaps most telling is the thermal emanations 
from the earth: not only is the heat evolved about twenty times 
larger than its ``supposed source'' from radioactive material, based 
upon the amount of helium effluxed, but this helium also contains 
$^{3}$He (not a component of radioactive decay from U and Th). 
Furthermore, there is evidence that at least some of the large scale 
magma deposits had their origin in surface-derived material, not from 
deep in the mantle. Thermal energy generation in the earth is discussed in another paper \cite{ref11}.
\par Finally, we should mention that if our tresino picture applied 
to dark matter is correct \cite{ref18}, it shows that the dark matter - the 
material filling most of the universe -- is composed of well-known  
entities - electrons and hydrogen nuclei.
\\


\begin{thebibliography}{99}
\bibitem{ref1} J. R. Reitz and F. J. Mayer, J. Math. Phys. \bf{41}\rm, 4572 (2000).
\bibitem{ref2} L. I. Schiff, \emph{Quantum Mechanics}, Second Edition, (McGraw-Hill, New York, NY, 1955). 
\bibitem{ref3} A. Messiah,\, \emph{Quantum Mechanics}-Vol.II, (John Wiley and Sons, 
New York, NY, 1962). See also, A. O. Barut and A. J. Bracken, Phys. 
Rev D, \bf {24}\rm, \, 3333 (1981). 
\bibitem{ref4} J. C. Slater, \emph{Quantum Theory of Atomic 
Structure}, pp189-192,(McGraw-Hill Book Co., New York, NY 1960), also 
Schiff, ibid, p283.
\bibitem{ref5} W. R. Smythe, \emph{Static and Dynamic Electricity}, 270-271, 
(McGraw-Hill Book Co., New York 1950).
\bibitem{ref6} M. Fleishmann and S. Pons, J. Electroanal. Chem. \bf{261}\rm,  301 (1989).
\bibitem{ref7} R. Notoya, Fusion Technology \bf 24\rm, 202 (1993).
\bibitem{ref8} R. Mills and K. Kneizys, Fusion Technology \bf 20 \rm, 65 (1991). 
\bibitem{ref9} J. A. Peacock, \emph{Cosmological Physics}, (Cambridge University Press,
Cambridge, UK, 1999) p. 384.
\bibitem{ref10} E. R. Oxburgh and R. K. O'Nions, Science \bf{237}\rm, 1583 (1987). 
\bibitem{ref11} F. J. Mayer and J. R. Reitz, ``Thermal Energy Generation in the Earth", to be published.
\bibitem{ref12} G. R. Satchler,\,\emph {Introduction to Nuclear Reactions}\rm, 
(Oxford University Press, New York, NY, 1990) pg. 67. Also see, Z. E. 
Switkowski, R. M. Wieland, and A. Winther, Phys. Rev. Lett. \bf{33} 
\rm, 840 (1974) for sub-barrier neutron transfer reactions.
\bibitem{ref13} C. G. Beaudette, \emph{Excess Heat: Why Cold Fusion Research Prevailed}, 
(Oak Grove Press, South Bristol, ME, 2000).
\bibitem{ref14} E. Storms, ``A Critical Evaluation of the Pons-Fleishmann Effect: Parts 
I and II'', Infinite Energy \bf{6}\rm (31), pg.10,
and \bf{6}\rm (32), pg.52 (2000), published by New Energy Foundation Inc. (Concord, NH). 
\bibitem{ref15}  R.A.Rice, Y. E. Kim, M. Rabinowitz and A.L. Zubarev,  ``Comments on Exotic Chemistry Models and Deep Dirac States for Cold Fusion",  Proc. International Conference Cold Fusion \em{4}\rm, 4-1- 4-7 (1993).
\bibitem{ref16} T. N. Claytor, D. G. Tuggle, H. O. Menlove, P. A. Seeger, W. R. Doty and 
R. K. Rohwer, ibid [14], p. 467.
\bibitem{ref17} W. B. Clarke, B. M. Oliver, M. C. McKubre, F. L. Tanzella, P. 
Tripodi, Fusion Science and Technology \bf{40}\rm, 152 (2001), and references therein.
\bibitem{ref18} F. J. Mayer and J. R. Reitz, ``Compton Composites Late in the Early Universe", to be published.


\end{thebibliography}
\end{document}